\documentstyle[12pt,epsf]{article}
\newcommand{\be}{\begin{equation}}
\newcommand{\ee}{\end{equation}}
\newcommand{\al}{\mbox{$\alpha$}}

\newcommand{\bi}[1]{\bibitem{#1}}
\newcommand{\fr}[2]{\frac{#1}{#2}}

\newcommand{\Tr}{\mbox{Tr}}

\begin{document}
\begin{titlepage}
\rightline{\vbox{\halign{&#\hfil\cr
&UQAM-PHE-98/01\cr
&\today\cr}}}
\vspace{0.5in}
\begin{center}

\Large\bf  Up-Down Unification just above the supersymmetric threshold.
\\
\medskip
\vskip0.5in

\normalsize {{\bf C. Hamzaoui}$^{\rm a}$\footnote{hamzaoui@mercure.phy.uqam.ca} and 
{\bf M. Pospelov}$^{\rm a,b}$}\footnote{pospelov@mercure.phy.uqam.ca} 
\smallskip
\medskip

{ \sl $^a$
D\'{e}partement de Physique, Universit\'{e} du Qu\'{e}bec a 
Montr\'{e}al\\ 
C.P. 8888, Succ. Centre-Ville, Montr\'{e}al, Qu\'{e}bec, 
Canada, H3C 3P8\\
$^b$Budker Institute of Nuclear Physics, Novosibirsk, 
630090, Russia} 
\smallskip
\end{center}
\vskip1.0in

\noindent
{\large\bf Abstract}
\smallskip\newline
Large corrections to the quark mass matrices at the supersymmetric threshold allow the theory to have identical Yukawa matrices in the superpotential. We demonstrate that Up-Down unification can take place in a moderate quark-squark alignment scenario with an average squark mass of the order 1 TeV and with $\tan\beta>15$. 

\end{titlepage}

\section{Introduction}
The observed hierarchical patterns in the masses of quarks and leptons and in the Kobayashi-Maskawa mixing suggest the existence of new physics beyond the standard model, perhaps in the form of new symmetries. Neither the character of these symmetries nor the scale of the new physics are understood so far. Various theoretical attempts have been made to construct a realistic explanation for the observed mass spectrum and reproduce experimentally observed values as close as possible. Some of them exploit horizontal symmetries \cite{FN,LNS}, gauged or nongauged, anomalous symmetries \cite{AS}, possible compositness \cite{K}, etc. 

The mass problem in supersymmetry has attracted serious theoretical attention in the last few years. Although very helpful for solving some of the shortcomings of the standard model, the supersymmetric framework by itself does not explain the observed patterns displayed by fermion masses. Yet it brings new phenomena in flavor physics related with the possible mismatch between quark and squark mass matrices. This mismatch, if significant, may induce unacceptable contribution to the neutral kaon mixing and therefore is constrained if the masses of the scalar quarks are not very far from the electroweak scale.

Another theoretical goal, closely related to the mass problem, is the possibility to reduce the number of the free parameters in the fundamental theory in comparison with that of the standard model. The supersymmetric GUT theories offer this possibility \cite{R}. The scale of the unification for Yukawa couplings in this case is believed to coincide with the scale of the gauge unification and therefore is very high, $10^{16}$GeV or so. 

In this letter, we ask how low the scale of the flavor unification could be. We  attempt to construct a model with a minimal number of free parameters in the superpotential. This is a model with identical Up and Down quark Yukawa matrices in the superpotential and general form of the soft-breaking parameters, not constrained by usual assumption of universality and proportionality. The Kobayashi-Maskawa mixing itself in this situation originates from the supersymmetric threshold corrections. We show that at the present stage of knowledge about supersymmetry breaking parameters this possibility is not excluded. 

There are several possible physical motivations related to  the $U$-$D$ unification. For example, it can be a consequence of a horizontal symmetry responsible for the generation of the Yukawa couplings which "does not feel hypercharge", i.e. which does not distinguish between $H_1$ and $H_2$, $U$ and $D$ superfields and therefore can only generate identical $Y_u$ and $Y_d$.  Another example is the case of the supersymmetric left-right theory based on the $SU(3)\times SU(2)_L\times SU(2)_R\times U(1)_{B-L}$ group \cite{MR} where the unification of $U$ and $D$ right-handed superfields is protected by an extra gauge symmetry. 

\section{U-D unification at low energy}
The standard superpotential of the minimal supersymmetric standard model (MSSM),
\be
W=\epsilon_{ij}[-Q^iH_2^j{\bf Y}_uU+Q^iH_1^j{\bf Y}_dD+
L^iH_1^j{\bf Y}_eE+\mu H_1^iH_2^j],
\label{eq:w}
\ee
contains the same number of free dimensionless parameters as the Yukawa sector of the standard model.

In addition, in the soft-breaking sector there are other couplings which have a potential influence on flavor physics. Among different scalar masses, the soft-breaking sector has the squark mass terms
\be
 \tilde{U}^{\dagger}{\bf M}^2_{U}\tilde U
+\tilde{D}^{\dagger}{\bf M}^2_{D}\tilde D+\tilde{Q}^{\dagger}{\bf M}^2_{Q}\tilde Q;
\ee
and the trilinear terms
\be
\epsilon_{ij}\left(-\tilde{Q}^iH_2^j{\bf A}_u\tilde{U}+
\tilde{Q}^iH_1^j{\bf A}_d\tilde{D}\right)~+H.c.;
\label{eq:A}
\ee
as the possible sources of flavor transitions. 

Counting all free parameters in the model, one comes to a huge number 105 \cite{D}.
This enormous number of free parameters originates mainly from the soft-breaking sector and cannot be reduced {\em a priori}, without knowledge of the ways the supersymmetry breaking occurs. It is customary to assume, at the scale of the breaking, the following, very restrictive conditions are fulfilled:
\begin{eqnarray}
\label{eq:deg}
{\bf M}_Q^2= m_Q^2{\bf 1};\;\;{\bf M}_D^2= m_D^2{\bf 1};
\;\;{\bf M}_U^2= m_U^2{\bf 1}\;\;\;\mbox{"degeneracy"}\\
{\bf A}_u= A_u{\bf Y}_u;\;\;{\bf A}_d= A_d{ \bf Y}_d\;\;\;\mbox{"proportionality"},
\label{eq:prop}
\end{eqnarray}
and similarly for leptons. These conditions, if held, ensure that the physics of flavor comes entirely from the superpotential. But it might not necessarily be the case. For example, these conditions are not held in superstring inspired models (See Ref. \cite{DKL} and references therein). Neither are they in the simplest flavor models operating with horizontal symmetries \cite{NS}.

To this end, it is interesting to abandon strict conditions (\ref{eq:deg})-(\ref{eq:prop}) in the soft-breaking sector and to explore the possibility of having a smaller number of free parameters in the superpotential. As an ultimate example of this, let us analyze the theory with the low-energy  unification of Up and Down Yukawa matrices. Instead of working with the superpotential of the form (\ref{eq:w})
with two independent matrices ${\bf Y}_u$ and ${\bf Y}_d$, we consider the theory with ${\bf Y}_u\equiv {\bf Y}_d$ so that the superpotenial can be written in the following compact form:
\be
W=\epsilon^{ij}\epsilon^{kl}[Q^i\Phi^{jk}{\bf Y }Q_R^l + \fr{1}{2}\mu \Phi^{jk}\Phi^{il}]
\label{eq:w1}
\ee
where $Q_R=(U,D)^T$ and $\Phi=(H_1;H_2)$. From here and below we consistently omit the leptonic part. At the same time we assume similar $U$-$D$ unification in the soft-breaking sector as well:
\be
...+ \tilde{Q}^{\dagger}{\bf M}^2_{L}\tilde Q~
+\tilde{Q}_R^{\dagger}{\bf M}^2_{R}\tilde Q_R~+
\epsilon^{ij}\epsilon^{kl}\tilde Q^i\Phi^{jk}
{\bf A }\tilde Q_R^l + H.c...
\ee
At the scale of the breaking, matrix ${\bf Y }$ can be chosen in the diagonal form and it will not develop any off-diagonal elements in the course of renormalization down to supersymmetric threshold scale. It means that, at the tree level plus logarithmic renormalization, $M_u\sim M_d$ and all Kobayashi-Maskawa mixing angles are zero. This might be considered, with a certain degree of optimism, as a good zeroth-ordrer approximation to the realistic mass matrices and mixing. In this case the observed mixing angles and masses come from the supersymmetric threshold corrections. These corrections induce additional terms in the Yukawa interaction containing the conjugated Higgs fields, $H_1^*$, $H_2^*$, and in our case $\Phi^*$. 
As a result, below the threshold, the effective interaction of fermions with Higgs doublets can be written in the form
\be
{\cal L}_{eff}=\bar Q {\bf Y}_1\tau_2\Phi\tau_2 Q_R + \bar Q {\bf Y}_2  \Phi^* Q_R+...
\label{eq:yuk}
\ee
The ellipses stands here for the possible terms of bigger dimension which also may influence the fermion spectrum (for example, $M_{sq}^{-2}\bar Q_R {\bf Y}_1\tau_2\Phi
\tau_2 Q\mbox{Tr}(\Phi^\dagger\Phi) $). In Fig. 1 we list the diagrams which contribute to the matrices ${\bf Y}_1$ and ${\bf Y}_2$. 

To some extent, one can view this unification as being inspired by supersymmetric left-right models where the relevant part of the superpotential is usually written in the following form:
\be
W=Q{\bf Y }_1\tau_2\Phi_1 \tau_2Q_R + Q{\bf Y }_2\tau_2\Phi_2\tau_2 Q_R + \sum_{i,j=1,2}\mu_{ij} \Tr \tau_2\Phi_i\tau_2\Phi^T_j
\label{eq:wlr}
\ee
If the analysis of the threshold correction can generate realistic ${\bf M}_u$ and ${\bf M}_d$
then one can eliminate one of the bidoublets, reducing (\ref{eq:wlr}) to the more economic form (\ref{eq:w1}).

The supersymmetric threshold corrections to the mass matrices and mixing have been considered before in a number of papers \cite{TC}. In most of these analysis, the conditions (\ref{eq:deg})-(\ref{eq:prop}) were imposed on the soft-breaking sector. In some specific variants of the unified models \cite{Hempfl}, however, the departure from these conditions can reproduce masses of first generation and the Cabbibo angle. For the recent discussion of the radiative mechanism for fermion masses in the presence of chiral symmetry violating soft-breaking terms see also Ref. \cite{BFPT}.

The typical size of the corrections to the mass of $b$-quark when the mass of the gluino is equal to the masses of squarks is of the order of $m_b$ itself, $\Delta m_b/m_b\sim 0.4 \mu/m_{squark}$. The ratio $\mu/m_{squark}$ is presumably of the order 1 but can be larger. For $m_{squark}\sim 1$TeV, $\mu/m_{sq}$ can be as large as $\sqrt{2}m_{sq}/v\sim 5$. In this case $m_b$, as well as the other quark masses and mixing angles, can be completely of radiative origin. In the following we  consider the possibility of the low-energy unification of Yukawa couplings, not confining our analysis to the case of $\mu\simeq m_{squark}$ and not specifying the details of a particular model which stays behind the origin of hierarchy. In particular, we have to check if the following three conditions are satisfied:
\begin{enumerate}
\item The matrix $\bf A$ is consistent with scale independent constraints from the absence of color breaking minima.
\item Radiatively generated masses and mixing angles correspond to the observed hierarchy. 
\item The predictions for FCNC are acceptable. 
\end{enumerate}

\section{Phenomenological consequences of U-D unification}

It is  convenient to choose the basis in which matrix ${\bf Y}$ is diagonal. In this case, superpotential contains only 3 dimensionless parameters in the quark sector
\be
{\bf Y}= \mbox{diag}(y_1;\,y_2;\,y_3)
\label{eq:diag}
\ee
and $y_3$ basically coincides with the top quark Yukawa coupling, $y_3 \simeq m_t\sqrt{2}/v\simeq 1$. 

The value of $\tan\beta\equiv v_u/v_d$ has to be large but it is not fixed to $m_t/m_b$ due to the substantial renormalization of $m_b$. Therefore, to $O(\tan^{-1}\beta)$ accuracy we can adopt the following approximation for the squark mass matrices:
\begin{eqnarray}
M_{\tilde u}^2 = \left(\begin{array}{cc}
{\bf M}^2_{L}+\fr{v^2}{2}{\bf Y}^{\dagger}{\bf Y}&
\fr{v}{\sqrt{2}}{\bf A}^{\dagger}\\
\noalign{\vskip10pt}
\fr{v}{\sqrt{2}}{\bf A} &
{\bf M}^2_{R}+\fr{v^2}{2}{\bf Y}{\bf Y}^{\dagger}
\end{array} \right)
\label{eq:Mu}
\end{eqnarray}
\begin{eqnarray}
M_{\tilde d}^2 = \left(\begin{array}{cc}
{\bf M}^2_{L}+\fr{v^2}{2}{\bf Y}^{\dagger}{\bf Y}&
\fr{v}{\sqrt{2}}\mu^* {\bf Y}^{\dagger}\\
\noalign{\vskip10pt}
\fr{v}{\sqrt{2}}\mu {\bf Y}&
{\bf M}^2_{R}+\fr{v^2}{2}{\bf Y}{\bf Y}^{\dagger}
\end{array} \right)
\label{eq:Md}
\end{eqnarray}

The general formulae for the mass matrices, tree level plus radiative corrections, can be presented in the following form:
\begin{eqnarray}
\fr{\sqrt{2}}{v}{\bf M}_{{\bf u}ij}={\bf Y}_{ij} + \fr{2\al_sm_\lambda}{3\pi}\int\fr{d^4p}{(2\pi)^4
(p^2-m_\lambda^2)}\left[\fr{1}{p^2-{\bf M}_L^2}{\bf A}\fr{1}{p^2-{\bf M}_R^2}\right]_{ij}+...\nonumber\\
\fr{\sqrt{2}}{v}{\bf M}_{{\bf d}ij}=\fr{{\bf Y}_{ij}}{\tan\beta}+\fr{2\al_s\mu m_\lambda}{3\pi}\int\fr{d^4p}{(2\pi)^4
(p^2-m_\lambda^2)}\left[\fr{1}{p^2-{\bf M}_L^2}{\bf Y}\fr{1}{p^2-{\bf M}_R^2}\right]_{ij}+ ...
\label{eq:mast}
\end{eqnarray}
Here, ellipses stands for the chargino and neutralino corrections and next order corrections in $v^2/m^2$ which we neglect at the moment.

The form of the matrix $A$ suggested by the absence of dangerous directions in the field space along which color breaking minima can emerge is the following \cite{CD}:
\be
\bf{A}=\left(\begin{array}{ccc} 0&0&A_{13}\\
0&0&A_{23}\\A_{31}&A_{32}&A_{33}
\end{array} \right)
\ee
The same form is suggested by the analysis of the FCNC processes \cite{Ch}. Restricting our analysis only to this form of ${\bf A}$ matrix, we satisfy the first condition mentioned in the previous section automatically. 

Let us consider the corrections to the mass of the b-quark.  Neglecting for a moment all flavor changing effects in the squark sector, we can identify $m_b$ with the $({\bf M_d})_{33}$ matrix element of the ${\bf M_d}$ and write the relation among observable $m_b$ taken at the scale of the squark mass, $\tan\beta$ and dimensionless ratios $\mu/m_{sq}$, $x=m_\lambda/m_{sq}$. For simplicity we assume also the approximate equality of the left- and right-handed squark masses. 
\begin{eqnarray}
\fr{2m_b^2}{v^2}=|y_3|^2\left|\fr{1}{\tan\beta}+ \fr{2\al_s}{3\pi}e^{i\phi}\fr{|\mu|}{m_{sq}}F(m_\lambda/m_{sq})\right|^2
\\\nonumber
F(x)=\fr{x}{1-x^2}+\fr{x^3\ln{x^2}}{(1-x^2)^2}
\end{eqnarray}
Taking $y_3\sim 1$, $m_b(1TeV)\sim 2.8 GeV$ and $F=F_{max}=F(2.1)=0.57$, we plot the allowed values of $\tan\beta$ and $|\mu|$ in Fig. 2. The phase $\phi$ between tree-level contribution and radiative corrections to $m_b$ is unknown \footnote{Possible constraints on $\phi$ from the limits on the neutron EDM ar relaxed when squark mass $\sim$ 1 TeV.} which leads to a certain allowed area on $\tan\beta$ - $|\mu|$ plane. The case of relatively low $\tan\beta$ corresponds to the destructive interference between the tree level contribution and the loop correction. With $\tan\beta\sim 15$, it corresponds to the 80\% mutual cancellation between tree-level value and radiative correction which we take as maximally allowed degree of the fine tuning. 

The necessity for the off-diagonal entries in the squark mass-matrices to be nonzero comes from two reasons. First, they can be the only source for the off-diagonal mass matrices leading to nonvanishing Kobayashi-Maskawa mixing angles. Second, their existence leads to  the substantial renormalization of $u,\,d,\, s,\, c$ quark masses which is needed to account for the relations $m_s/m_c>m_b/m_t;\;  m_d/m_u\gg m_b/m_t$.  The 100\% renormalization of the charm mass, for example, may result from the combination of the flavor changing entries in ${\bf A}$ and $ {\bf M}_R^2$, and so on. At the same time, it is preferable to keep the off-diagonal elements of the squark mass matrices at the lowest possible level to avoid large FCNC contributions from the box diagrams. Treating these flavor transitions as the mass insertions, i.e. assuming that they are small in some sense, we arrive at the following set of the order-of-magnitude relations connecting observable masses and mixing angles to flavor structure of the soft-breaking sector:  
\begin{eqnarray}
\fr{\sqrt{2}}{v}\Delta m_c\sim \eta_c
\fr{A_{23}({\bf M}_R^2)_{23}}{m^3_{sq}}\nonumber\\
\fr{\sqrt{2}}{v}\Delta m_s\sim \eta_s
\fr{({\bf M}_L^2)_{23}({\bf M}_R^2)_{23}}{m^4_{sq}}\nonumber\\
\fr{\sqrt{2}}{v}\Delta m_d\sim \eta_d
\fr{({\bf M}_L^2)_{13}({\bf M}_R^2)_{13}}{m^4_{sq}}\\
\theta_{23}\sim\fr{({\bf M}_L^2)_{23}}{m^2_{sq}}\nonumber\\
\theta_{13}\sim\fr{({\bf M}_L^2)_{13}}{m^2_{sq}}\nonumber
\label{eq:om}
\end{eqnarray}
Here $m_{sq}$ is the average squark mass and numerical coefficients $\eta_i$ represent loop factors. For our estimates we take $\eta_i$ to be of the order of  $\eta_b\sim \sqrt{2}m_b/v$. As to the mixing angle between first and second generation, it can be generated either in the down sector \cite{Hempfl}, or in the up sector through the correction to the matrix element $({\bf M_u})_{12}\sim \fr{\sqrt{2}}{v}\eta A_{13}({\bf M}_R^2)_{23}/m^3_{sq}$.

If the squark masses were diagonal, one would observe an approximate relation $m_b/m_t\simeq m_s/m_c$ which is violated in reality. There are several possible ways to avoid this problem depending on which part of the allowed values of $\mu$ - $\tan\beta$ plane we choose. If $\tan\beta$ is in the neighbourhood of $m_t/m_b$, it is preferable to have $y_2v/\sqrt{2}\simeq m_c$ and the strange quark mass being completely of radiative origin.
The latter condition requires $({\bf M}_L^2)_{23}/m^2_{sq}\sim \lambda^2$ and
$({\bf M}_R^2)_{23}/m^2_{sq}\sim 1$. For lower values of $\tan\beta$, when the tree-level contribution to the mass of bottom quark is compensated by loop corrections to give observable $m_b$, the element $({\bf M}_R^2)_{23}$ can be made smaller and $y_2v/\sqrt{2}\simeq m_s$. 

Even with the assumptions of the low energy U-D unification, the number of free parameters is large enough to reproduce the observed masses and mixings. Combining together relations (\ref{eq:om}), we write the phenomenologically acceptable form of the squark mass matrices which can produce correct flavor physics through the loop mechanism:
\be
{\bf M}_L^2\sim m^2\left(\begin{array}{ccc} 1&\lambda^4&\lambda^2\\
\lambda^4&1&\lambda^2\\\lambda^2&\lambda^2&1
\end{array} \right);\;\;\;
{\bf M}_R^2\sim m^2\left(\begin{array}{ccc} 1&\lambda&\lambda\\
\lambda&1&1\\\lambda&1&1
\end{array} \right).
\label{eq:align}
\ee
Every entry in  (\ref{eq:align}) denotes an order of magnitude of corresponding matrix element in terms of the power of Wolfenstein parameter $\lambda$. The squark masses chosen in the form (\ref{eq:align}), plugged in the general formula (\ref{eq:mast}), reproduce correctly the hierarchy among quark masses and mixing angles and therefore satisfies the second condition formulated at the end of the previous section. $({\bf M}_{L(R)}^2)_{12}$ is not fixed by (16) and for its value we take $({\bf M}_{L(R)}^2)_{12}\sim({\bf M}_{L(R)}^2)_{13}({\bf M}_{L(R)}^2)_{23}/m^2_{sq}$. 

Similar  forms of the squark mass matrices can appear in supersymmetric theories with horizontal symmetries responsible for mass hierarchy \cite{NS} (quark-squark alignement). We can invert the set of argument and conclude that in the quark-squark alignment picture with large $\tan\beta$ regime and $\mu m_\lambda\sim m_{sq}^2$ the radiative corrections to the mass matrices and mixings are important. At the same time, the bare superpotential of the model  can be of the form (\ref{eq:w1}), i.e. simpler than that of the conventional MSSM. 

We turn now to the "effective supersymmetry" picture \cite{ES} which has much less degrees of freedom and where flavor physics can be formulated in the more definite way. In the squark mass matrices, diagonalized by unitary transformations $U$ and $V$,
\be
{\bf M}_L^2=U^\dagger \left(\begin{array}{ccc} m^2_1&0&0\\
0&m^2_2&0\\0&0&m^2_3
\end{array} \right)U;\;\;
{\bf M}_R^2=V^\dagger \left(\begin{array}{ccc} m'^2_1&0&0\\
0&m'^2_2&0\\0&0&m'^2_3
\end{array} \right)V,
\ee
the eigenvalues $m^2_1$, $m^2_2$, $m'^2_1$, $m'^2_2$ are taken to be in the multi-TeV scale and eventually decoupled from the rest of particles. At the same time, the squark from the third generation is believed to be not heavier than $1$ TeV and weakly coupled to the first and second generations of quarks to  avoid the excessive fine-tuning in the radiative corrections to the Higgs potential and suppress FCNC contribution to the Kaon mixing. The advantage of this approach in our case is that the loop integrals
 in (\ref{eq:mast}) can be parametrized by one number $\eta$. Moreover, Up quark Yukawa matrix keeps its nearly diagonal form and the Kobayashi-Maskawa mixing results entirely from ${\bf M_d}$. Introducing the notations   
\be
U^*_{13}= \omega, \,\, U^*_{23}= \rho,\,\, U^*_{33}\simeq 1,\,\, V_{13}= \gamma, V_{23}= \delta,\,\, V_{33}= \sqrt{1-|\gamma|^2-|\delta|^2},
\ee
we write the resulting form of the mass matrix:
\be
\fr{\sqrt{2}}{v}{\bf M_d}= \eta\left(\begin{array}{ccccc} \omega\gamma& &
\omega\delta& 
&\omega\sqrt{1-|\gamma|^2-|\delta|^2}\\ & & & & \\ \rho\gamma& 
&\fr{y_2}{\eta\tan\beta}+\rho\delta& &\rho
\sqrt{1-|\gamma|^2-|\delta|^2}\\& & & & \\
\gamma& &\delta& &
\fr{y_3}{\eta\tan\beta}+
\sqrt{1-|\gamma|^2-|\delta|^2}
\end{array} \right)
\label{eq:esmm}
\ee 
Here we take into account that $y_3\simeq 1$ and $\rho,~\omega\ll 1$.

Using this form of the mass matrix, we calculate $H_d=\fr{2}{v^2}{\bf M_d}{\bf M_d}^\dagger$, its eigenvalues and Kobayashi-Maskawa matrix which diagonalizes $H_d$. The determinant of $H_d$ has a simple form:
\be
det H_d=|\eta|^6|\omega|^2|\gamma|^2
\fr{|y_2|^2}{|x|^4},
\label{eq:det}
\ee
where $x=\eta\tan\beta$. The presence of $y_2$ in  the 
$det H_d$ suggests to choose $y_2v/\sqrt{2}\simeq m_s$.
This requires the partial compensation of tree level and radiative corrections in $m_b$ and as a result one has $x\simeq -1$ and $\delta \sim O(\lambda)$. 

The analysis of $|V_{us}|$ yields even the smaller value for the parameter $\gamma$. It turns out that to sufficient accuracy one has
\be
|V_{us}|^2  = \left(\fr{m_d^2}{m_s^2}\right)\left(\fr{2m_s^2}{v^2|y_2|^2}
\right)
\fr{|\delta|^2+|\gamma|^2}{|\gamma|^2}
\ee
and therefore $\gamma \sim O(\lambda^2)$.
At the same time 
\be
|V_{ub}|=|\omega|\fr{\left|1+\fr{1}{x}\sqrt{1-|\delta|^2}\right|}
{\left|1+\fr{1}{x}\sqrt{1-|\delta|^2}\right|^2
+|\delta|^2}
\ee
which makes $|\omega|$ be of the order of $\lambda^3$. A similar relation for $|V_{cb}|$ shows that $\rho\sim O(\lambda^2)$. Thus the analysis of the mass matrices generated by the loop with third generation of squarks inside gives $\rho, ~\omega,~\gamma, ~\delta \ll 1$ in consistency with similar requirement imposed by the absence of large FCNC contributions.

The FCNC processes in the neutral Kaon sector arises both at the tree level and at one-loop level. When the FCNC contribution of the box diagrams generated by the mass matrices (\ref{eq:align}) is considered,  the constraints \cite{FCNC}  implies that the lowest possible value for the squark mass has to be of the order $1$ TeV. Similar constraints arise from the analysis of $b\rightarrow s\gamma$ process \cite{bsg}. This can be still viewed as an acceptable value for the scalar quark mass scale. 

Perhaps the most serious consequence of the low-energy unification of Up and Down Yukawa matrices and radiative mechanism for Kobayashi-Maskawa mixing is the appearence of the FCNC transitions mediated by $H_d$ field. Since the resulting Yukawa interaction resembles that of a generic left-right model with the right-hadned mixing angles not smaller than left-handed ones, $|V_{Rij}|>|V_{Lij}|$, we can use the limits on the FCNC Higgs masses obtained in \cite{G}:
\be
M_A> 10\mbox{TeV} 
\ee
We would like to emphasize here that in the generic MSSM with large $\tan\beta$ the tree level FCNC induced due to the threshold corrections to the mass matrices may exceed contributions from box diagrams  and therefore should be properly accounted for in the general analysis  \cite{FCNC}.
 
\section{Conclusions}

The supersymmetric mass problem has one interesting aspect, not always properly emphasized. With the general flavor-changing soft-breaking terms, it is hard to interpret the dimensionless coefficient in the superpotential in terms of the observed fermionic masses and mixing angles. This opens up a possibility of a supersymmetric theory with much less number of the free parameters in the superpotential than it is assumed in the conventional MSSM. We have shown the phenomenological possibility of the low-energy $U$-$D$ unification in the supersymmetric models. The Kobayashi-Maskawa mixing in this case is the result of the supersymmetric threshold corrections. The analysis of these corrections shows that $\tan\beta$ is not  fixed by the requirement of the unification. $U$-$D$ unification allows it to be in a rather large range $15\leq \tan\beta\leq \infty$. 

It was shown in Refs. \cite{LNS} that the scale of the physics responsible for the flavor hierarchy can be as low as few TeV scale. Our conclusion is that the condition of the unification of all Yukawa couplings does not require to raise this scale. The scale of the unification, i.e. supersymmetric threshold in the case considered, can be as low as 1TeV without causing unacceptable mixing in the neutral kaon sector. 

An obvious advantage of the "effective supersymmetry" approach in the connection with the low energy $U-D$ unification is that the resulting mass matrix (\ref{eq:esmm}) can be parametrized by a few number of parameters which are the mixing angles between third generation squark and quark fields. At the same time all loop integrals are expressed by one number $\eta$ and this  significantly simplifies the analysis. The condition imposed on this mass matrix to reproduce observable masses and mixings automatically leads to the smallness of the mixing angles between third generation squark and first and second generation of quarks, which coincides with the similar requirements imposed by the absence of FCNC. 

The possibility of the $U$-$D$ unification at low energies with the radiative mechanism for Kobayashi-Maskawa mixing has an interesting application to the  left-right supersymmetric models. It allows one to reduce the number of Higgs bidoublets, making it more similar to the conventional (nonsupersymmetric) left-right models. Unfortunately, in the case of the manifest left-right symmetry, where left-handed and right-handed squark masses are equal, the radiatively induced fermion masses and mixings do not correspond to the observables. 
\smallskip

\mbox{\hspace{-0.5cm}}{\large {\bf Acknowledgements}} \newline This work is supported in part by N.S.E.R.C. of Canada.

\newpage
\large{\bf Figure captions }\normalsize

Figure 1.  The diagrams generating threshold corrections for ${\bf M_u}$  and  ${\bf M_d}$.

Figure 2.  Allowed area on the $|\mu|-\tan\beta$ plane in the case of maximal threshold corrections.

\vspace{1cm}
\begin{figure}[hbtp]
\begin{center}
\mbox{\epsfxsize=144mm\epsffile{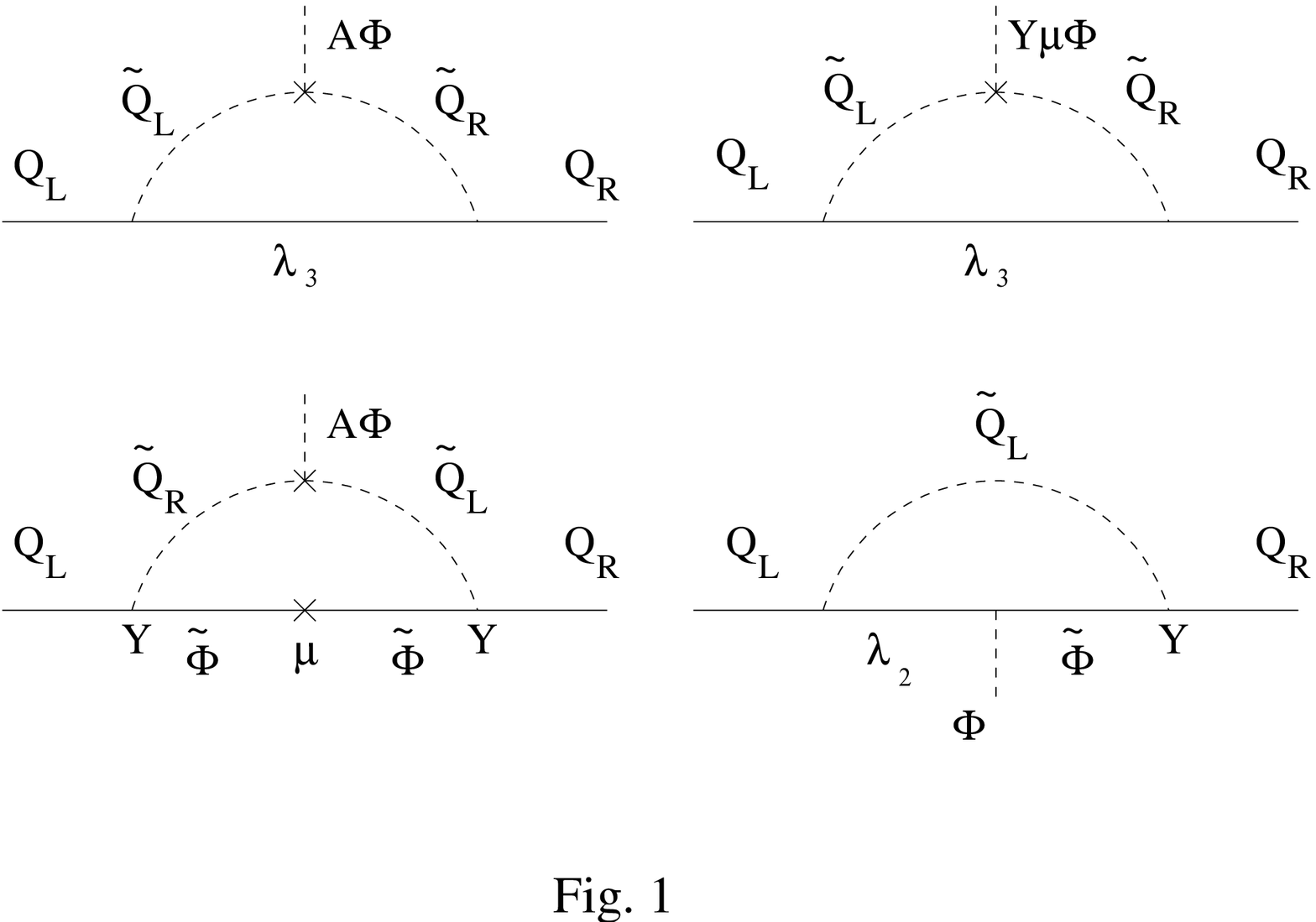}}
\end{center}
\end{figure}
\newpage
\begin{figure}[hbtp]
\begin{center}
\mbox{\epsfxsize=144mm\epsffile{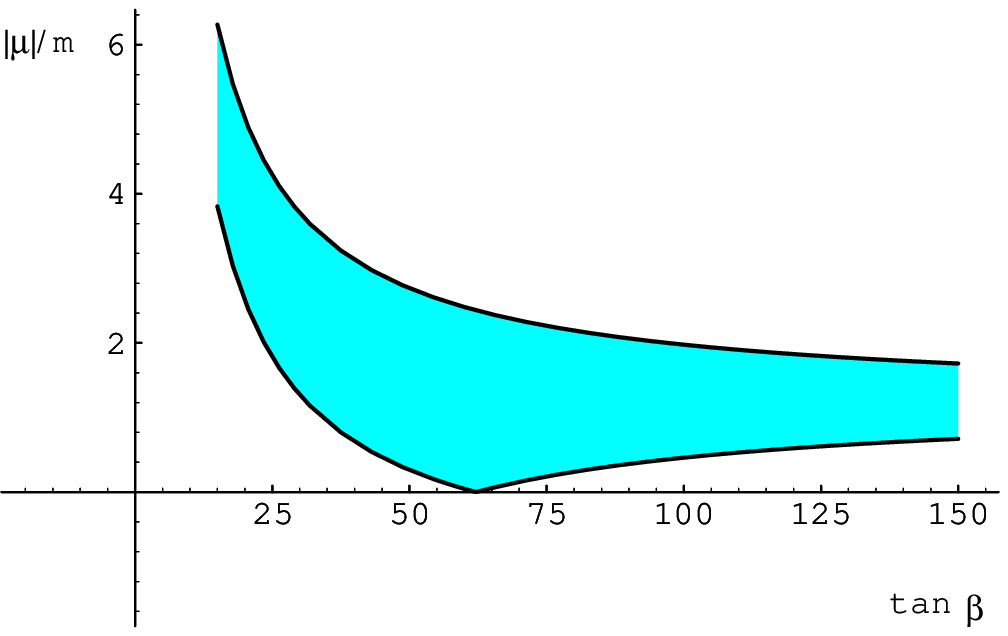}}
\end{center}
\end{figure}
\begin{center}\large Figure 2\end{center}

\end{document}